\documentclass[12pt,english,longbibliography,aps,nofootinbib,preprint]{revtex4-1}
\usepackage{hyperref}
\usepackage{setspace}
\usepackage{mathptmx}
\usepackage{graphicx}

\usepackage[T1]{fontenc}
\usepackage[latin9]{inputenc}
\usepackage[letterpaper]{geometry}
\usepackage{tikz}
\usepackage{parskip}
\usepackage{booktabs}
\usepackage{amsmath}
\usepackage{amssymb}
\usepackage{amsthm}
\usepackage{cancel}
\usepackage{babel}
\usepackage{xcolor}
\usepackage{mathtools}
\newtheorem{theorem}{Theorem}[section]
\newtheorem{corollary}[theorem]{Corollary}
\newtheorem{lemma}[theorem]{Lemma}

\theoremstyle{definition}

\begin{document}

\title{Absence of $\sigma$-criticality in Short-Range Spin Glasses}
\begin{abstract}
The stability of ground states in the Edwards-Anderson Ising spin glass in dimensions two and higher against perturbations of a single coupling can be quantified through the ``flexibility'' of that coupling in a particular ground state.  The property of $\sigma$-criticality is one in which varying a single coupling can alter the flexibilities of a (strictly) positive density of couplings in an infinite system, and its presence provides a sufficient condition for the presence of multiple ground state pairs.   In a previous paper it was shown for the EA spin glass that $\sigma$-criticality of the first kind, in which the critical droplet of a bond has a positive-density interface with the ground state, was ruled out in any dimension.  The remaining possibility, called $\sigma$-criticality of the second kind, is one in which a bond lies in the critical droplet boundaries of a positive density of other bonds.  We show here that this kind of $\sigma$-criticality is also absent, ruling out altogether the presence of $\sigma$-criticality in the EA spin glass. While this doesn't resolve the question of multiplicity of ground state pairs, it removes a sufficient condition for its presence. We then propose a rigorous necessary condition for ground state pair multiplicity.
\end{abstract}
\author{C.M.~Newman}
\affiliation{Courant Institute of Mathematical Sciences, New York University, New York, NY 10012 USA}
\affiliation{NYU-ECNU Institute of Mathematical Sciences at NYU Shanghai, 3663 Zhongshan Road North, Shanghai, 200062, China}
\author{D.L.~Stein}
\affiliation{Department of Physics and Courant Institute of Mathematical Sciences, New York University, New York, NY 10012 USA}
\affiliation{NYU-ECNU Institutes of Physics and Mathematical Sciences at NYU Shanghai, 3663 Zhongshan Road North, Shanghai, 200062, China}
\affiliation{Santa Fe Institute, 1399 Hyde Park Rd., Santa Fe, NM USA 87501}

\maketitle

\section{Introduction}
\label{sec:intro}

How is a spin glass ground state affected by a change in the coupling realization? This question of stability of both finite- and infinite-volume ground states in classical short-range Ising spin glasses is intimately connected to fundamental thermodynamic properties, including multiplicity of distinct ground states, low-temperature thermodynamics, and nonequilibrium dynamics at zero and low temperatures~\cite{BM87,FH88b,NS99a,NS99c,NS2D00,NS2D01,NS03a,KB05,KK07,NS08,Chatterjee09,ADNS10,ANS19,JRY21,ANS21,NS22,Chatterjee23,NS24c,Nussinov24,NS26}.  Two approaches toward answering this question have been studied.  The first~\cite{BM87,FH88b,KB05,KK07,Chatterjee09,ANS19} is known as disorder chaos, which focuses on the behavior of a ground state when each of the coupling values has changed independently by a small random amount.  The second approach~\cite{NS2D00,NS2D01,ADNS10,ANS19,ANS21,NS22,Chatterjee23,NS24c,Nussinov24,NS26}, with which we will concern ourselves in this paper, focuses on changes induced in a ground state when a {\it single\/}  coupling value is changed --- in a direction to destabilize the ground state --- by an~arbitrary amount.  The robustness or sensitivity of spin ground states to finite coupling changes provides a means to distinguish among various proposed scenarios of the spin glass phase~\cite{NS22,NS26}, as well as a possible avenue toward addressing the long-unresolved question regarding multiplicity of ground states in finite dimensions.

This paper addresses the property of $\sigma$-criticality, defined in~\cite{NS22}. A spin glass ground state is $\sigma$-critical if a finite change in a single coupling can induce a global decrease in stability of the ground state.  To make this notion precise, the concept of flexibility was introduced in~\cite{NS2D00,NS2D01}, and a definition appears below in Sect.~\ref{sec:gs}.  Roughly speaking, the flexibility of a single coupling describes the magnitude of the change needed in its value to destabilize the ground state, i.e., to induce a droplet flip leading to a new ground state.  We denote this flipped droplet as the {\it critical droplet\/} (also defined in Sect.~\ref{sec:gs}) of the corresponding bond.  The size of the critical droplet flip induced by the coupling change is of particular importance, in terms of both the number of spins contained in the droplet bulk and the number of bonds contained in its boundary.  One can then define the average flexibility of a ground state, namely the average of the flexibility of all its couplings in a given coupling realization.  This then gives a precise quantification of the stability of a ground state: the lower the average flexibility, the less stable the ground state.  

From this perspective, a $\sigma$-critical ground state is the least stable of all spin glass ground states, in the sense that altering the value of a single coupling reduces the average flexibility of the entire ground state in the infinite-volume limit. There are two types of $\sigma$-criticality, defined for infinite-volume ground states.  $\sigma$-criticality of the first kind occurs when the critical droplet of an arbitrarily chosen bond has a positive probability of having a boundary comprising a positive density of bonds. $\sigma$-criticality of the second kind occurs when an arbitrarily chosen bond has a positive probability of belonging to the critical droplet boundary of a positive density of other bonds.  In~\cite{NS22} it was shown that ground state flexibility is directly connected to the question of ground state multiplicity; in particular, Theorem 9.3 of~\cite{NS22} shows that the presence of either kind of $\sigma$-criticality is a sufficient (but not necessary) condition for the spin glass to possess multiple distinct ground states (modulo a global spin flip).

In~\cite{NS26} it was shown that $\sigma$-criticality of the first kind cannot occur (with nonzero probability in the coupling realization) in short-range Ising spin glasses in finite dimensions whose couplings are chosen from a continuous distribution with finite variance.  In this paper we similarly rule out $\sigma$-criticality of the second kind, showing that this sufficient condition for ground state multiplicity cannot occur.

The organization of this paper is as follows.  Sect.~\ref{sec:gs} provides a review of basic definitions and relevant features of spin glass ground states, critical droplets and their flexibilities. In Sect.~\ref{sec:sigma2} we show that if a bond has the property of $\sigma$-criticality of the second kind at a given coupling realization $J$, then it retains that property over a nonzero range of its coupling values with all other couplings fixed. In Sect.~\ref{sec:meta} we prove that $\sigma$-criticality of the second kind cannot occur (with nonzero probability in the couplings) in short-range spin glasses under the same conditions~\cite{NS26} as those for which $\sigma$-criticality of the first kind was ruled out. Finally, in Sect.~\ref{sec:necessary} we derive, using this general approach, a rigorous {\it necessary\/} condition for observable ground state multiplicity to occur.

\section{Ground states, critical droplets, and flexibility}
\label{sec:gs}

In this section we review relevant results obtained in previous work.  For a more comprehensive treatment, we refer the reader to~\cite{NS22,NS26}.

We study the behavior of ground states in the Edwards-Anderson (EA) classical Ising spin glass model~\cite{EA75} in zero magnetic field on the $d$-dimensional cubic lattice $\mathbb{Z}^d$. Its Hamiltonian is
\begin{equation}
\label{eq:EA}
{\cal H}_J=-\sum_{<x,y>} J_{xy} \sigma_x\sigma_y 
\end{equation}
where $\sigma_x=\pm 1$ is the Ising spin at site $x$ and $\langle x,y\rangle$ denotes an edge (or equivalently, bond) between nearest-neighbor sites $x$ and $y$; we denote the set of all edges as $\mathbb{E}^d$. Each edge $\langle x,y\rangle\in\mathbb{E}^d$ is assigned a coupling $J_{xy}$, where the $J_{xy}$'s are independent, identically distributed random variables chosen from a distribution $\nu(dJ_{xy})$.  Our requirements on $\nu$ are that it be continuous (with no atoms), distributed symmetrically about zero, with finite variance  and full support on ${\mathbb R}$.

As in~\cite{NS26}, our focus is on finite- and infinite-volume ground states of the EA~spin glass in finite dimensions $d\ge 2$.  Let $\Lambda_L$ be a cube of side~$L$ centered at the origin; a finite-volume ground state~$\sigma_L$ is then the lowest-energy spin configuration in~$\Lambda_L$ subject to a specified boundary condition (which we usually take to be periodic). An infinite-volume ground state~$\sigma$ can be defined either as any convergent $L\to\infty$ limit of a subsequence of $\sigma_L$'s, or equivalently, a spin configuration $\sigma$ on all of $\mathbb{Z}^d$ defined by the condition that its energy cannot be lowered by flipping any {\it finite\/} subset of spins.  This can be expressed mathematically as
\begin{equation}
\label{eq:gs}
E_{S}\equiv\sum_{\langle x,y\rangle^*\in{S}}J_{xy}\sigma_x\sigma_y\ >0 
\end{equation}
where $\langle x,y\rangle^*$ denotes the dual of the edge $\langle x,y\rangle$ and $S$ is any closed surface (or contour in two dimensions) in the dual lattice.  Hereafter we denote by $b^*_{xy}$ the dual of bond $b_{xy}\in\mathbb{E}^d$. 

We turn next to critical droplets, which were introduced in~\cite{NS2D00,NS2D01}.  Because critical droplets and their properties have been extensively discussed in earlier papers (see especially~\cite{NS22} and~\cite{NS26}, but also~\cite{ANS19,ANS21,Nussinov24}), we give a very brief summary here, listing definitions and results which will be needed below, and refer the reader to to the above papers for details.

In what follows we will use the notation $J_b$ to refer to the (fixed) coupling value assigned to the bond $b$ in the coupling realization $J$. We will often vary the value of a single coupling belonging to a bond $b$ while holding all other couplings fixed;  we will use the notation $J(b)$ to refer to this running coupling value.

\medskip

{\bf Definition.}  [Newman-Stein~{\cite{NS2D00}]}.  Consider the finite-volume GSP $\sigma_L$ for the EA Hamiltonian~(\ref{eq:EA}).  Choose a bond $b$ and consider all surfaces in the dual edge lattice $E_L^*$ which include the dual edge $b^*$ and which partition the spins in $\Lambda_L$ into two disjoint sets. The energies of these surfaces are given by~Eq.~(\ref{eq:gs}) and so are all positive.  Because (by continuity of the coupling distribution) there is zero probability that any two such surfaces have equal energy, there must exist one of {\it least\/} energy in $\sigma_L$. We call this surface the {\it critical droplet boundary\/} of $b$ in $\sigma_L$ and denote it by $\partial D(b,\sigma_L)$. We further define the {\it critical droplet\/} of $b$ in $\sigma_L$ as the set of spins $D(b,\sigma_L)$ enclosed by $\partial D(b,\sigma_L)$.

\medskip

{\bf Definition}. [Newman-Stein~{\cite{NS2D00}]}. The {\it critical value\/} in $\sigma_L$ of the coupling assigned to the bond $b$ is denoted $J_c^{\sigma_L}(b)$ (or simply $J_c^{\sigma_L}$ if the bond in question is unambiguous) and is the value of $J(b)$ where $E\bigl(D(b,{\sigma_L})\bigr)=0$, while all other couplings in $J$ are held fixed.

\medskip 

{\bf Definition}. [Newman-Stein~{\cite{NS2D00}]}. Suppose $J_b=K$ in $J$. We define the {\it flexibility\/}~of $J_b$ at that particular value to be $f(J_b,{\sigma_L})=\vert K-J_c^{\sigma_L}(b)\vert$.  Similarly, if the value of that coupling varies, its flexibility is defined to be $f(J(b),{\sigma_L})=\vert J(b)-J_c^{\sigma_L}(b)\vert$.

\medskip

\begin{lemma}
\label{lem:infinite}
{\rm [Newman-Stein~\cite{NS22}, Lemma 3.1].  Critical droplets, critical values of couplings, and flexibilities remain well-defined and their properties preserved in the infinite-volume limit.}
\end{lemma}

\medskip

{\bf Remark.} The proof of Lemma~2.1 involves construction of the excitation metastate~\cite{NS2D00,NS2D01,ADNS10,ANS19}; flexibility values and their associated critical droplets converge together at least along a subsequence of finite volumes used in the construction of the metastate.

\medskip

Before proceeding, we note that critical droplets (denoted $D(b,\sigma)$) and their boundaries (denoted $\partial D(b,\sigma)$) are defined with respect to bonds (because critical droplets are geometric objects that are independent of the bond's coupling value). Flexibilities are defined with respect to couplings because they're the distance between a coupling at a particular value and its critical point. Similarly, critical droplet energies (denoted by $E\bigl(D(b,\sigma)\bigr)$) are defined with respect to couplings.

In what follows we list a number of previously obtained results which will be needed in the proofs below.  As noted above the couplings are chosen from a continuous distribution $\nu(dJ)$ with zero mean and finite variance.

\medskip

\begin{lemma} 
\label{lem:share}
{\rm [Newman-Stein~\cite{NS22}, Lemma 2.5]. Consider two distinct edges $b$ and $b'$ and an infinite-volume ground state $\sigma$. (a) If $f(J_{b'},\sigma) > f(J_{b},\sigma)$, then $b'^*$ cannot belong to $\partial D(b,\sigma)$, while $b^*$ may or may not belong to~${\partial D}(b',\sigma)$.  Conversely, if $b^*\in\partial D(b',\sigma)$ and $b'^*\notin\partial D(b,\sigma)$, then $f(J_{b'},\sigma) > f(J_b,\sigma)$. (b) If $b$ and $b'$ share the same critical droplet, then w.p.~1 $b^*\in\partial D(b',\sigma)$ and $b'^*\in\partial D(b,\sigma)$ (the converse is true as well).  If $b^*$ and $b'^*$ share the same critical droplet, then by~(\ref{eq:flex}), $J_b$ and $J_{b'}$ have equal flexibilities.}
\end{lemma}

\medskip

\begin{lemma} 
\label{lem:flexlower}
{\rm [Newman-Stein~\cite{NS26}, Lemma~2.3]. If the flexibility of any given coupling is lowered (by changing its coupling value) but remains positive in $\sigma$, the flexibility of any other coupling in $\sigma$ is either also lowered (by up to the same amount) or else remains unchanged.  Similarly, if the flexibility of any coupling is increased, then the flexibility of any other edge in $\sigma$ is either also raised (by up to the same amount) or else remains unchanged. The same applies to the energy change of all droplets in the system, whether critical or not.}
\end{lemma}

\medskip

\begin{lemma}
\label{lem:remain}
{\rm [Newman-Stein~\cite{NS22}, Lemma 6.2].  Suppose a bond $b_0$ with coupling value $J_0$ in $J$ and critical value $J_c^\sigma(b_0)$ belongs to the critical droplet boundary $\partial D(b_1,\sigma)$ of a different bond $b_1$. Then $b_0$ will remain in $\partial D(b_1,\sigma)$ for the entire range of coupling values between $J_0$ and $J_c^\sigma(b_0)$.}
\end{lemma}

\medskip

\begin{corollary}
\label{cor:flexener}
{\rm With $f(J(b),\sigma)$ denoting the flexibility of the coupling $J(b)$ in the GSP~$\sigma$, and $E\bigl(D(b,\sigma)\bigr)$ denoting the energy of its critical droplet using~(\ref{eq:gs}) (with $S$ being the critical droplet boundary $\partial D(b,\sigma)$), we have~\cite{NS22}}
\begin{equation}
\label{eq:flex}
f\bigl(J(b),\sigma\bigr)=E\bigl(D(b,\sigma)\bigr)\, .
\end{equation}
\end{corollary}

\medskip

\begin{theorem}
\label{thm:cont}
{\rm The flexibility of any coupling either remains unchanged or else changes continuously as either it or any other coupling passes through its critical value in either direction.}
\end{theorem}

\medskip

{\bf Remark.} This is a corollary of Lemma 3.2~of~\cite{NS26}.

\medskip

\begin{lemma}
\label{lem:cont2}
{\rm If a coupling $J(b_0)$ is varied without crossing its critical value then the energy of any (fixed) droplet either varies continuously with $J(b_0)$ (if $b_0^*$ lies in its boundary) or else remains unchanged (if $b_0^*$ does not lie in its boundary).}
\end{lemma}

\medskip

{\bf Proof.} This follows immediately from the definition of the energy of a droplet.

\medskip

\begin{lemma}
\label{lem:zero}
{\rm [Newman-Stein~\cite{NS22}].  There is zero probability in $J$ that a coupling $J_b$ is exactly at its critical value (and therefore has zero flexibility) in a GSP~$\sigma$.}
\end{lemma}

{\bf Remark.}  This never appeared separately as a lemma in earlier papers, but was shown to be true in~\cite{NS22} (see the Remark just above Eq. (4) in~\cite{NS22}).

\bigskip

The following two results were proved using a periodic boundary condition metastate $\kappa_J$ but are valid for any translation-covariant metastate $\eta_J$; discussions can be found in~\cite{NS22,NRS23b,NS26}.  In the rest of the paper we will use the more general metastate $\eta_J$.

\bigskip

\begin{theorem}
\label{thm:flexconst}
{\rm [Newman-Stein~\cite{NS22}, Lemma~7.2].  For fixed $J$, let $P_J(f,\sigma)$ denote the (empirical) probability distribution over all edges of the flexibilities $f$ in the ground state $\sigma$, and $P_J(f)=\langle P_J(f,\sigma)\rangle_{\eta_J}$ be the metastate average of $P_J(f,\sigma)$ over the ground states~$\sigma$ in the support of $\eta_J$. Then $P_J(f)$ is a.s.~constant with respect to $J$.}
\end{theorem}

\medskip

\begin{theorem}
\label{thm:nosf}
{\rm [Newman-Stein~\cite{NS26}, Theorem~4.13]. Choose a GSP~$\sigma$ from $\eta_J$. Then with $(J,\kappa_J)$-probability one, every bond $b$ has a critical droplet whose boundary is either finite (enclosing a finite set of spins) or else has zero density in~$\mathbb{E}^d$.}
\end{theorem}

\section{$\sigma 2$ bonds and ground states}
\label{sec:sigma2}

All critical droplets are defined for infinite volume~$\mathbb{Z}^d$ using constructions described in~\cite{NS2D01,NS22}.  Throughout this paper ``positive'' means ``strictly positive''.

{\bf Definition.}   If, for some open interval of coupling values $J(b_{xy})$, $b^*_{xy}$ belongs to the critical droplet boundaries of a positive density (in $\mathbb{E}^d$) of bonds in a GSP~$\sigma$,  we say it exhibits $\sigma$-criticality of the second kind for that $\sigma$~\cite{NS22}.  We will refer to such a bond as a~$\sigma 2$ bond.

\medskip

{\bf Remark.}   The $\sigma 2$ property is the property of a given bond in a specified GSP~$\sigma$. For convenience, when we say that a bond~$b_0$ has the $\sigma 2$ property, or simply that $b_0$ is $\sigma 2$, we mean that it has the $\sigma 2$ property for some range of coupling values (which are ground-state-dependent) in a particular, previously specified ground state $\sigma$.  We will also say that a GSP $\sigma$ is $\sigma 2$ (for a particular coupling realization $J$) if its configuration is such that a positive density of bonds has the $\sigma 2$ property.

\medskip

\begin{theorem}
\label{thm:posdens}
{\rm A GSP drawn from~$\eta_J$ either has no $\sigma 2$~bonds or else a positive density of them.}
\end{theorem}

\medskip

{\bf Proof.} This is proved by the same reasoning as in the proof of the theorem~in~\cite{NS01c}.~$\diamond$

\medskip

In the remainder of this section we will investigate how the $\sigma 2$ property of a bond is affected upon changing the value of its coupling with all others held fixed.  We show that if a bond $b_0$ has the $\sigma 2$ property at any single coupling value $J_0>J_c^\sigma(b_0)$, then it retains the property for the entire interval $(J_c,J_0]$. The analogous conclusion holds for $J_0<J_c^\sigma(b_0)$.

\medskip

\begin{lemma}
\label{lem:cc}
{\rm Consider a fixed coupling realization $J$ and a corresponding GSP $\sigma$ with the $\sigma 2$ property at that $J$. Choose a bond $b_0$ having the $\sigma 2$ property, and suppose its associated coupling $J(b_0)$ has the value $J_0> J_c$ in $J$. Then $b_0$ retains the $\sigma 2$ property for all $J(b_0)$ in the entire interval $(J_c,J_0]$ with all other couplings held fixed.  If $J_0<J_c$, the analogous conclusion holds throughout $[J_0,J_c)$.}
\end{lemma}

\medskip

{\bf Proof.}  For notational convenience we will hereafter refer to $J_c(b_0,\sigma)$ as $J_c$.  Throughout the paper we hold all coupling values fixed with the exception of that associated with $b_0$, denoting its coupling's running value by $J(b_0)$.

Suppose that $b_0$ has the $\sigma 2$ property in $\sigma$ at $J_0$ and set $a=J_0-J_c>0$.  Choose any bond~$b_1$ with $b^*_0$ in its critical droplet boundary at $J(b_0)=J_0$; by assumption, these bonds constitute a positive density of all bonds.  By Lemma~\ref{lem:remain} it follows that $b^*_0$ remains in the critical droplet boundary of $b_1$ throughout the interval $J(b_0)=(J_c,J_c+a]$.  Because this is true for all such bonds with $b_0^*$ in their critical droplet boundaries, and given that such bonds constitute a positive density of all bonds, $b_0$ retains the $\sigma 2$ property throughout the interval $(J_c, J_c+a]$.  If $J_0<J_c$, the same argument holds throughout $[J_0,J_c)$.~$\diamond$  

\medskip

Lemma~\ref{lem:cc} shows that if any bond $b_0$ is $\sigma 2$ at any value of $J(b_0)\ne J_c$ then it retains the $\sigma 2$ property over an interval of values of $J(b_0)$ between that value and $J_c$. 
It does not guarantee that $b_0$ retains the $\sigma 2$ property if $J(b_0)$ crosses $J_c$.  Suppose the $\sigma 2$ bond $b_0$ has coupling value $J_0$ in coupling realization $J$. A corresponding GSP can then be one of three types: (i) $J_0>J_c$ and $b_0$ loses its $\sigma 2$ property when $J(b_0)<J_c$; (ii) $J_0<J_c$ and $b_0$ loses its $\sigma 2$ property when $J(b_0)>J_c$; and (iii) $b_0$ retains its $\sigma 2$ property on both sides of $J_c$. (As always, $J(b_0)$ is the only variable coupling; all other couplings are always held fixed.)  From here on we will refer to the corresponding GSP's as type~(i), type~(ii), and type~(iii), respectively.

The interval of coupling values $J(b_0)$ for which $b_0$ has the $\sigma 2$ property in {\it any\/} ground state must be finite; there is an upper bound to its length, as we now explain.

\medskip

{\bf Definition}.  We say that a bond $b_{xy}$ (or its coupling $J(b_{xy})$) is {\it supersatisfied\/} in some fixed coupling realization $J$ if it is satisfied in every GSP.

It is not hard to see that a value of $\vert J(b_{xy})\vert$ above which $b_{xy}$ must be supersatisfied is 
\begin{equation}
\label{eq:ss}
\vert J(b_{xy})\vert\ge J^{\rm ub}(b_{xy})={\min}\Bigl(\sum_{z\ne y\atop |z-x|=1}\vert J_{xz}\vert,\sum_{u\ne x\atop |y-u|=1}\vert J_{uy}\vert\Bigr)
\end{equation}
so the maximum length of the interval outside of which any bond must be supersatisfied is $2\vert J^{\rm ub}\vert$; however, some bonds could be supersatisfied outside a smaller interval (an example is given in~\cite{NS26}).  We will denote the interval outside of which $b_0$ is supersatisfied as $(J_{\rm lower},J_{\rm upper})$. The range of values of $J(b_0)$ for which $b_0$ is $\sigma 2$ (as well as all critical values of $J(b_0)$ over all ground states) must therefore lie inside the bounds implied by~(\ref{eq:ss}).  We will use $J^\sigma_c(b_0)\in(J_{\rm min},J_{\rm max})\subseteq(J_{\rm lower},J_{\rm upper})$ to denote the maximal range of critical values of $J(b_0)$ within which $b_0$ is $\sigma 2$ in {\it some\/} GSP~$\sigma$; i.e., if $J_c^\sigma(b_0)$ falls outside $(J_{\rm min},J_{\rm max})$, $b_0$ is not $\sigma 2$ for that $\sigma$.  Of course $b_0$ might not be $\sigma 2$ for every $\sigma$ with $J_c^\sigma(b_0)$ within this interval, but it cannot be $\sigma 2$ for a $\sigma$ with $J_c^\sigma(b_0)$ outside it.  

\section{Metastates and type-2 criticality}
\label{sec:meta}  

In this section we study the presence or absence of $\sigma 2$ criticality in ground states in the support of $\eta_J$, any zero-temperature translation-covariant EA metastate (discussions of metastates and their properties can be found in~\cite{NS22,NRS23b,NS26}).  An often-used metastate belonging to this class is a periodic boundary condition metastate, generated using an infinite deterministic sequence of volumes with periodic boundary conditions.

Before proceeding, we prove a lemma that will be needed in what follows.

\medskip

\begin{lemma}
\label{lem:concave}
{\rm  On an interval of the running coupling $t = J(b_0)$ over which a GSP~$\sigma$ does not flip, each bond flexibility $f(b,\sigma,t)$ is concave in $t$.}
\end{lemma}

\medskip

{\bf Proof.}  In finite volumes this follows because flexibility is the minimum of the energies of all competing droplet boundaries through $b$, with each energy being affine in $t$ while the spin configuration stays fixed.  By Lemma~\ref{lem:infinite}, the finite-volume critical values converge as do the flexibility functions along the same (excitation metastate) construction at the coupling values being compared; in particular, concavity is preserved by pointwise limits of functions. On a compact interval avoiding the limiting critical value, sufficiently large finite volumes have no change in ground state, so their concavity inequalities are preserved in the infinite-volume limit.~$\diamond$

\subsection{Positive weights}
\label{subsec:positive}

\begin{theorem}
\label{thm:finite}
{\rm  The metastate $\eta_J$ cannot be supported only on a finite set of GSP's with each having positive weight in~$\eta_J$ and at least one having bonds with the $\sigma 2$ property.}
\end{theorem}

\medskip

{\bf Proof.} Let $N$ denote the number of GSP's in the support of $\eta_J$, with $1\le N<\infty$. Suppose first that $N=1$ with the corresponding GSP~$\sigma$ having the $\sigma 2$ property.  We choose a bond $b_0$ with the $\sigma 2$ property and suppose that in the initial coupling realization~$J$ its coupling value $J(b_0)=J_0$; by Lemma~\ref{lem:zero}, $J_0\ne J_c$.  By Lemmas~\ref{lem:flexlower} and~\ref{lem:cc}, and the definition of $\sigma 2$-criticality, varying $J(b_0)$ toward $J_c$ (but without crossing it, and holding all other couplings fixed) will change the average flexibility of $\sigma$, contradicting Theorem~\ref{thm:flexconst}.

We next turn to the case where $2\le N<\infty$.  By assumption there exists a bond $b_0$ that has the $\sigma 2$ property in $1\le n\le N$ of the ground state pairs. We label this subset of ground state pairs as $\sigma_1, \sigma_2, \ldots \sigma_n$ with $J_{c1}\ge J_{c2}\ge \ldots\ge J_{cn}$, where $J_{ci}$ is the critical value of $J(b_0)$ in ground state pair~$\sigma_i$. Label the remaining GSPs $\sigma_{n+1},\ldots,\sigma_N$.

Given that there are finitely many GSP's, the initial coupling is almost surely at none of their critical values. Let $J(b_0)=J_0$ in the coupling realization $J$. Set  $0<h<\min_{1\le j\le N}\vert J_0-J_c^{\sigma_j}(b_0)\vert$, and choose a GSP~$\sigma^*$ in which $b_0$ has the $\sigma 2$ property at $J_0$. If $J_0>J_c^{\sigma^*}(b_0)$, set $c=J_0-h$ and $d=J_0$.
If $J_0<J_c^{\sigma^*}(b_0)$, set $c=J_0$ and $d=J_0+h$. In either case, as $J(b_0)$ moves toward $J_c^{\sigma^*}$ within $[c,d]$, no GSP undergoes a critical droplet flip and (by Lemma~\ref{lem:remain}) $b_0$ preserves its $\sigma 2$ property in $\sigma^*$.

Let $B$ be the positive-density set of bonds whose critical-droplet boundaries contain $b_0^*$ in $\sigma^*$ at $J_0$. By Lemma~\ref{lem:remain} ,
this inclusion persists throughout $[c,d]$. By Corollary~\ref{cor:flexener} and the affine droplet boundary energy dependence, each bond in $B$ has flexibility decreasing affinely, with slope of magnitude one, as $J(b_0)$ moves from $J_0$ toward the other endpoint. Moreover, under the coupling-covariance of the metastate, all GSP weights remain fixed.  

Consider the bounded test function on flexibilities
\begin{equation}
\label{eq:test}
\phi(f)=1-e^{-\lambda f},\qquad\lambda>0\, .
\end{equation}
Since $\phi$ is increasing and concave as a function of flexibility, Lemma~\ref{lem:concave} implies that $\phi(f(b,\sigma,t))$ is concave in $t$.   Thus the spatial and metastate averages of $\phi$ are concave as well.  

Using the above definitions, the strict concavity of $\phi$ (as explained in the following paragraphs) gives a strict concavity inequality 
\begin{equation}
\label{eq:conc1}
\phi\left(f\Big(\frac{c+d}{2}\Big)\right)>\frac{\phi\left(f(c)\right)+\phi\left(f(d)\right)}{2}
\end{equation}
for this population.  

Let $t=J(b_0)$ and define
\begin{equation}
\label{eq:A}
A_\phi(t) =\sum_{i=1}^{N}W_i\Bigl(\lim_{L\to\infty}\frac{1}{\vert E_L\vert} \sum_{b\in E_L}\phi\Big(f(b,\sigma_i,t)\Big)\Bigr)
\end{equation}
where $W_i$ is the metastate weight of $\sigma_i$.  Because $\phi(f)$ is a bounded continuous function of the flexibility, Theorem~\ref{thm:flexconst} implies that $A_\phi(t)=\int \phi(f)P_t(df)$ is a.s.~constant, where $P_t(df)$ denotes the metastate flexibility distribution under the running coupling $J(b_0)=t$.  On the other hand, the strict concavity property requires that $A_\phi\bigl((1-\theta)t_0+\theta t_1\bigr) > (1-\theta)A_\phi(t_0)+\theta A_\phi(t_1)$ for suitable points in the interval. 

As an illustration, consider the case $J_0>J_c^{\sigma^*}(b_0)$, so that the perturbation lowers $J(b_0)$ from $d$ to $c$.  Let $b$ denote a bond with $\sigma 2$ activity in $(c,d)$ in the GSP~$\sigma^*$. If $f(b,\sigma^*,d)=x$ then $f(b,\sigma^*,c)=x-h$.  Because $f(b,\sigma^*)$ decreases affinely as $J(b_0)$ is lowered from $d$ to $c$, $f\left(\frac{c+d}{2}\right)=x-\frac{h}{2}$. In this case the difference between the two sides in~(\ref{eq:conc1}) is $e^{-\lambda(x-h/2)} \left[\cosh\left(\frac{\lambda h}{2}\right)-1\right]>0$. We will call this difference the ``concavity gap'' for $b$ in $\sigma^*$; note the concavity gap is positive whether flexibility increases or decreases.  Every other bond contribution in every supported GSP also has a nonnegative concavity gap. 

To ensure strict concavity after spatial averaging choose a sufficiently large finite $M>0$ and restrict to a positive density subset of the affected bonds whose couplings have flexibility less than or equal to $M$; i.e., this bounded-flexibility subset is selected using $f(b,\sigma^*,d)\le M$. The empirical flexibility distribution allows this restriction, and on that bounded range the strict concavity gap has a positive uniform lower bound for a fixed perturbation. Note that the strict inequality and its uniform bound also hold for any fixed $0<\theta<1$.  

Now fix $b_0$ and condition on all other couplings. By Fubini and Theorem~\ref{thm:flexconst}, for almost every such conditioning, $A_\varphi(t)$ equals its deterministic value on a full~$\nu$-measure set ${\mathcal G}$. Because $\nu$ has full support, ${\cal G}$ is dense. On the open interval under consideration, $A_\varphi$ is finite and concave, hence continuous. It must therefore be constant throughout that interval, contradicting the strict concavity established above.~$\diamond$

\medskip

\begin{theorem}
\label{thm:countable}
{\rm The metastate $\eta_J$ cannot be supported only on a countably infinite set of GSP's at least one of which has the $\sigma 2$ property.}
\end{theorem}

\medskip 

{\bf Proof.}  The proof is identical to that of Theorem~4.4 in~\cite{NS26}.~$\diamond$

\medskip

\begin{corollary}
\label{cor:mixed}
{\rm A GSP with positive weight in any zero-temperature metastate $\eta_J$ cannot have the $\sigma 2$ property.}
\end{corollary}

\medskip

{\bf Proof.} The only remaining case to be considered is a ``mixed'' scenario, where part of the support of $\eta_J$ is on GSP's with zero weight and part on GSP's with positive weight.  The method used in the proof of Theorem~4.4 in~\cite{NS26} applies here as well.~$\diamond$

We can therefore consider in what follows scenarios where $\eta_J$ is supported solely on GSP's with zero weight.

\subsection{Zero weights}
\label{subsec:zero}

Suppose now that $\eta_J$ is supported entirely on an uncountable infinity of GSP's each having zero $\eta_J$-measure, and suppose that the set of GSP's having the $\sigma 2$ property have positive $\eta_J$-measure. Because this set of GSP's is uncountable, whereas the set of edges in ${\mathbb E}^d$ is countable, there must exist a bond $b_0$ with the $\sigma 2$ property in a subset of GSP's with positive $\eta_J$-measure.  We denote by ${\cal A}$ the set of $\sigma$'s in which $b_0$ is $\sigma 2$. 

In what follows we will require a result from~\cite{NS26}, which places a bound on the range of values for which a $\sigma 2$ bond in a given ground state possesses the $\sigma 2$ property.  We start by defining (slightly differently from~\cite{NS26} where $a^+(b_0,\sigma)$ was positive) $a^+(b_0,\sigma)\ge 0$ to be the largest value for which $J(b_0)\in\Big(J_c^\sigma(b_0),J_c^\sigma(b_0)+a^+(b_0,\sigma)\Big)$ retains the $\sigma 2$ property (with $a^+(b_0,\sigma)=0$ indicating that for $J(b_0)>J_c^\sigma(b_0)$ the GSP loses the $\sigma 2$ property); similarly, $a^-(b_0,\sigma)\ge 0$ is defined as the largest value for which $J(b_0)\in\Big(J_c^\sigma(b_0)-a^-(b_0,\sigma),J_c^\sigma(b_0)\Big)$ retains the $\sigma 2$ property.  If $b_0$ is type~(i) in $\sigma$ (cf.~the discussion following the proof of~Lemma~\ref{lem:cc}), $a^+(b_0,\sigma)>0$ and $a^-(b_0,\sigma)=0$; if type~(ii), $a^+(b_0,\sigma)=0$ and $a^-(b_0,\sigma)>0$; and if type~(iii) $\sigma$, both $a^+(b_0,\sigma)>0$ and $a^-(b_0,\sigma)>0$.

\medskip

\begin{lemma}
\label{lem:H}
{\rm [Adapted from {\bf Newman-Stein~\cite{NS26}, Lemma 4.7}.] Choose an interval $(c,d)\subseteq(J_{\rm min},J_{\rm max})$ such that $p>0$, where $p$ is the metastate measure of $\sigma$'s with $J_c^\sigma\in(c,d)$. Suppose further that a positive fraction of these $\sigma$'s have $a^+(b_0,\sigma)>0$. Finally, let $H({\hat a}^+):=\{\sigma:\ \sigma\ {\rm has}\ a^+(b_0,\sigma)\ge{\hat a}^+>0\}$.   Then for any such $(c,d)$ there exists a $k>0$ and a ($k$-dependent) ${\hat a}^+>0$ such that at least a fraction $kp>0$ of $\sigma$'s with $J_c^\sigma\in(c,d)$ belongs to $H({\hat a}^+)$. The same result holds for $a^-(b_0,\sigma)$ with all quantities defined analogously.}
\end{lemma}

\medskip

{\bf Remark.} The same value of $k$ could (in principle) lead to different values of ${\hat a}^+$ and ${\hat a}^-$.

\medskip

{\bf Remark.} Lemma~4.7 of~\cite{NS26} mistakenly asserted that the result of the Lemma held for all $k\in(0,1]$, where in fact it holds only for some $k>0$. This does not affect any of the subsequent theorems or conclusions in that paper.

\medskip

{\bf Remark.}  Lemma~\ref{lem:H} does not exclude the possibility that a particular value of $J(b_0)$ may be a {\it null point\/}. A value $J^*$ of the coupling $J(b_0)$ is a null point if $J_{\rm lower}<J^*<J_{\rm upper}$ and one (or both) of the following two possibilities occur(s): a) there exists~$J_1<J^*$ such that $a^+(b_0,\sigma)\le \vert J^*-J_c^\sigma\vert$ for almost all $\sigma$ having $J_c^\sigma\in(J_1,J^*)$, and/or b) there exists~$J_2>J^*$ such that $a^-(b_0,\sigma)\le \vert J^*-J_c^\sigma\vert$ for almost all $\sigma$ having $J_c^\sigma\in(J^*,J_2)$.  If a) occurs at $J^*$, we will refer to it as ``left-null'', and if b) occurs at $J^*$, we refer to it as ``right-null''.  (For trivial reasons $J_{\rm upper}$ has the left-null property and $J_{\rm lower}$ has the right-null property, so the definition excludes these values given that the null property is of interest only if the corresponding coupling value lies strictly between $J_{\rm upper}$ and $J_{\rm lower}$.) Because the possible presence of null points does not alter any of the conclusions and at most requires only a slight modification of the proofs (see the Remark following Theorem~4.12 in~\cite{NS26}), we ignore the possible presence of null points in the proofs of the following theorems.

\medskip

Because $J(b_0)$ is the only coupling that will be varied, we hereafter denote its critical value in the GSP $\sigma$ by $J_c^\sigma$. By the discussion above, $J_c^\sigma\in(J_{\rm lower},J_{\rm upper})$ for all $\sigma$ in the support of $\eta_J$.  If $(J_{\rm min},J_{\rm max})\subset(J_{\rm lower},J_{\rm upper})$, then for $\sigma\in{\cal A}$ we need be concerned only with critical values $J_c^\sigma\in(J_{\rm min},J_{\rm max})$.  

In what follows we will let ${\cal I}^>$ denote the interval $(J_{\rm lower},J_{\rm upper})$ and ${\cal I}^<$ denote the interval $(J_{\rm min},J_{\rm max})$.  

\subsubsection{$J_{\rm max}<J_{\rm upper}$ {\rm or} $J_{\rm min}>J_{\rm lower}$}
\label{subsubsec:gap}

Suppose that $J_{\rm max}<J_{\rm upper}$ so that there is a gap between the upper ends of ${\cal I}^>$ and ${\cal I}^<$.  We first consider a scenario in which the $J_c^\sigma$'s corresponding to $\sigma$'s with the $\sigma 2$ property are dense in an interval $(c,J_{\rm max})$.

\medskip

\begin{theorem}
\label{thm:subsetdense}
{\rm Consider the event in which $J_{\rm max}<J_{\rm upper}$ and the set of $J_c^\sigma$'s with $\sigma\in{\cal A}$ are dense in the interval $(c,J_{\rm max})$ where $J_{\rm min}\le c<J_{\rm max}$ and with $\eta_J(\{\sigma: J_c^\sigma\in (v,w)\})>0$ for all  $(v,w)\subseteq(c,J_{\rm max})$, while some interval $(u,c)$ with $u<c$ is empty of $J_c^\sigma$'s with $\sigma$ having the $\sigma 2$ property.  This event occurs with zero $\nu$-probability.} 
\end{theorem}

\medskip

{\bf Proof.} All $\sigma$'s referred to in the proof have the $\sigma 2$ property.

If a fraction~one of $\sigma\in(c,J_{\rm max})$ are type~(i), then as $J(b_0)$ is lowered from a starting value just above $J_{\rm max}$ to a final value just above $c$, $\langle f\rangle_{\eta_J}$ must decrease by Lemma~\ref{lem:flexlower}, given that $a^-(b_0,\sigma) = 0$ for these states.   

If a fraction~one of $\sigma\in(c,J_{\rm max})$ are type~(ii), then using the same process of lowering $J(b_0)$ as above, the flexibility now increases in a positive-measure subset of these GSP's and remains unchanged in the others.  However, ground states with $J_c^\sigma\le c$ may contribute a decrease in the average flexibility if their $\sigma 2$ activity extends above $c$. To control this possible cancellation, let ${\mathcal B}$ denote the set $\{\sigma: J_c^\sigma\le c\}$ for which $b_0$ is type~(i) or~(iii) and $J_c^\sigma + a^+(b_0,\sigma)>c$.   Set $R={\rm ess}\ {\rm sup}_{\sigma\in{\mathcal B}}(J_c^\sigma+a^+(b_0,\sigma))$ where the essential supremum is with respect to $\eta_J$, and set $R=c$ if $\eta_J({\mathcal B})=0$.

Consider first the case where $R>J_{\rm max}$. Then there exists $\delta>0$ such that $\eta_J(\{\sigma\in{\mathcal B}: J_c^\sigma+a^+(b_0,\sigma)>J_{\rm max}+2\delta\})>0$. Lower $J(b_0)$ from just above $J_{\rm upper }$ to $J_{\rm max}+\delta$. Every state in this positive-measure subset is $\sigma 2$ throughout the common interval $(J_{\rm max} + \delta, J_{\rm max} + 2\delta)$, and its average flexibility decreases during the perturbation. The type-(ii) states with critical values in $(c,J_{\rm max})$ have no effect. The resulting decrease in the average flexibility contradicts Theorem~\ref{thm:flexconst}.

If $R<J_{\rm max}$, then lower $J(b_0)$ from just above $J_{\rm upper }$ to a value strictly between $R$ and $J_{\rm max}$; the lower population is inactive, while a positive-measure upper type-(ii) population contributes an increase to the average flexibility.

We now turn to the final case $R= J_{\rm max}$.   The procedure here is essentially identical to that used in Theorems~4.11 and~4.12 in~\cite{NS26}, which shows that the flexibility does not remain constant in this case for a suitably chosen interval of variation of $J(b_0)$.  In the current proof $J_{\rm max}$ corresponds to $J_{\rm upper}$ in the proofs of Theorems~4.11 and~4.12 in~\cite{NS26}.  We direct the reader to~\cite{NS26} for the full arguments.

This rules out the possibilities that $\sigma$'s with the $\sigma 2$ property and $J_c^\sigma\in(c,J_{\rm max})$ are all type~(i) or all type~(ii), leaving only the case where type~(iii) $\sigma$'s comprise a positive fraction of the GSP's with $J_c^\sigma\in(c,J_{\rm max})$.~\footnote{It could also happen that some bonds are type (i) and the remaining bonds are type~(ii) with no type~(iii)  bonds.  If the type~(i) and type~(ii) bonds are segregated, e.g. the bonds with larger $J_c$ are all type~(i) and those with smaller $J_c$ are all type~(ii), the same arguments as above go through.  If the two types of bonds are interspersed, then the argument used for type~(iii) bonds will work for this case as well.}  Suppose that $J_{\rm max}$ is not a ``left-null'' point as defined in the third Remark following Lemma~\ref{lem:H}.  Then 
\begin{equation}
\label{eq:H+}
\eta_J\!\left( \left\{\sigma:J_c^\sigma\in(c,J_{\max}),\ J_c^\sigma+a^+(\sigma)>J_{\max}\right\} \right)>0\, . 
\end{equation}
This implies the existence of $\delta > 0$ for which a positive-measure subset of type~(i) or type~(iii) GSP's is $\sigma 2$ throughout $(J_{\rm max}+\delta, J_{\rm max}+2\delta)$.
With all other couplings held fixed, set $J(b_0)=J_{\rm max}+2\delta$, and lower it to just above $J_{\rm max}+\delta$. This lowers in turn the average flexibility of $\eta_J$, leading to a contradiction with Theorem~\ref{thm:flexconst}.

A similar argument can be used for the case where $J_{\rm min}>J_{\rm lower}$.~$\diamond$

\medskip

We next turn to a scenario in which the $J_c^\sigma$'s are not dense in $(c,J_{\rm max})$. 

\medskip

\begin{theorem}        
\label{thm:subsetnondense}
{\rm Consider the situation in which $J_{\rm max}<J_{\rm upper}$ with the set of $J_c^\sigma$'s belonging to $\sigma\in{\mathcal A}$ not dense in any subinterval of $(c,J_{\rm max})$, but with the set within the entire interval~$(c,J_{\rm max})$ having positive $\eta_J$-measure. This situation occurs with zero $\nu$-probability.}
\end{theorem}

\medskip

{\bf Proof.}  The proof is similar to that of Theorem~\ref{thm:finite} with a few modifications.  Let $\widehat J$ denote the set of all couplings in $J$ except the coupling associated with $b_0$; also let $t=J(b_0)$, the running coupling value of $b_0$.  Let $\omega\in\Omega$ label a ground-state family $\sigma_\omega(t)$ (including its excitation data) as $t$ varies. Equip $\Omega$ with a probability measure $\mu_{\widehat J}$, independent of $t$, such that, for every measurable collection ${\mathcal B}$ of ground states, 
\begin{equation}
\label{eq:omega}
\mu_{\widehat J}(\omega:\sigma_\omega(t)\in {\mathcal B}) =\eta_{(\widehat J,t)}({\mathcal B}) \quad\text{for }\nu\text{-a.e.}\ t\, ;
\end{equation}
i.e, $\omega$ is the (coupling-dependent) family of spin configurations that follow from $\sigma$ as $J(b_0)$ is varied with all other couplings held fixed.  We will refer to $\omega$ as a {\it trajectory\/} of $\sigma$. 

Next let ${\mathcal M}=\{\omega: \left[ (J_c^\sigma(\omega)-a^-(\omega),J_c^\sigma(\omega)) \cup (J_c^\sigma(\omega),J_c^\sigma(\omega)+a^+(\omega)) \right]\cap (c,J_{\rm max})\ne\varnothing\}$. We will call the set of trajectories in ${\mathcal M}$ `active'.   The hypotheses of the theorem give 
\begin{equation}
\label{eq:mu}
\mu_{\widehat J} \Bigl({\omega\in\mathcal M: J_c^\sigma(\omega)\in(c,J_{\max})}\Bigr)>0\, ; 
\end{equation}
the critical values of this trajectory population satisfy the nondensity condition.

Now define $K=\overline{\{J_c^\sigma(\omega):\omega\in\mathcal M\}}$, where $\overline{\{J_c^\sigma(\omega):\omega\in\mathcal M\}}$ denotes the closure of $\{J_c^\sigma(\omega):\omega\in\mathcal M\}$. Every trajectory counted in~(\ref{eq:mu}) has a nonempty open activity interval inside $(c,J_{\max})$. Since $K$ has empty interior relative to $(c,J_{\max})$, this activity interval contains a nondegenerate rational segment, together with a neighborhood, disjoint from $K$. The corresponding countable collection of trajectory sets covers the positive-measure population in~(\ref{eq:mu}), so (at least) one member has positive measure.  Note that the hypothesis of a nondense set of critical values in the statement of the theorem applies to the critical values of the full active trajectory population inside $(c,J_{\max})$.

With these considerations, define the interval
\begin{equation}
\label{eq:uv}
[u,v]\subset (c,J_{\rm max})\setminus \overline{\{J_c^\sigma(\omega):\omega\in\mathcal M\}}\, .
\end{equation}
That is, the interval $[u,v]$ is a closed segment of coupling values strictly inside $(c,J_{\max})$, chosen so that no active trajectory has its critical value in or arbitrarily close to $[u,v]$ but a positive $\mu_{\widehat J}$-measure of trajectories is active throughout $[u,v]$.  Using the countability of rational segments, we can choose this segment so that a positive-measure set ${\mathcal S}\subset{\mathcal M}$ is active throughout a neighborhood of it, as discussed above~(\ref{eq:uv}).  

As in the proof of Theorem~\ref{thm:finite} we now use the bounded, increasing concave test function 
\begin{equation}
\label{eq:phi2}
\phi(f)=1-e^{-\lambda f}
\end{equation}
with $\lambda>0$ along with
\begin{equation}
\label{eq:phi2}
F_\omega(t)=\lim_{L\to\infty}\frac1{|E_L|} \sum_{b\in E_L}\phi(f_b(\omega,t))
\end{equation}
and
\begin{equation}
\label{eq:phi3}
A_\phi(t)=\int_\Omega F_\omega(t)\,\mu_{\widehat J}(d\omega)\, .
\end{equation}
For almost every conditioning on $\widehat J$, choose a countable dense subset of the full $\nu$-measure set of coupling values at which the empirical limits exist for $\mu_{\widehat J}$-a.e.~trajectory. On a single set of trajectories of full measure, the limits then exist at every point of this countable set. The uniform Lipschitz bound established in~(c) below extends these limits uniquely to all coupling values and implies convergence of the spatial averages there as well.

As $J(b_0)$ varies through $[u,v]$, there are three contributions to $\phi(f)$: 

\noindent a) active trajectories outside ${\mathcal S}$. The critical droplets of the corresponding $\sigma$'s do not flip as $J(b_0)$ varies, and Lemma~\ref{lem:concave} then gives nonnegative concavity gaps for these $\sigma$'s.

\medskip

(b) active trajectories in ${\mathcal S}$. For each such trajectory, select the positive-density set of bonds whose critical-droplet boundaries contain $b_0^*$ at the endpoint of $[u,v]$ farther from its critical value. Lemma~\ref{lem:remain} preserves this inclusion throughout the segment. By Corollary~\ref{cor:flexener} and the affine critical droplet boundary energy dependence, the flexibilities of these bonds vary affinely with slope of magnitude one throughout $[u,v]$. The same bounded-flexibility truncation as in Theorem~\ref{thm:finite} gives a strictly positive spatially averaged concavity gap.

\medskip

(c) inactive trajectories. For an inactive trajectory, consider two coupling values on the same side of its critical value. Any bond whose flexibility changes must have its critical-droplet boundary containing $b_0^*$ at the endpoint nearer the critical value. This follows from the finite-volume comparison of droplet energies and passes to the limit by the joint convergence in Lemma~\ref{lem:infinite}. Inactivity implies that these bonds have zero spatial density, so the spatial average of the bounded function $\phi$ is unchanged. Moreover, Lemma~\ref{lem:flexlower} and Theorem~\ref{thm:cont}  give $\vert f_b(\omega,t)-f_b(\omega,s)\vert\le |t-s|\), including across the critical value. Because $\phi'(f)=\lambda e^{-\lambda f}\le\lambda$ for $f\ge 0$, we have $\vert\phi(f_b(\omega,t))-\phi(f_b(\omega,s))\vert\le\lambda\vert t-s\vert$. The same bound holds for the finite spatial averages and hence for their limits: $\vert F_\omega(t)-F_\omega(s)\vert\le\lambda\vert t-s\vert$.  Thus these averages are continuous across the critical value. Then $F_\omega(t)$ is constant on $[u,v]$ for inactive trajectories, including those whose ground states undergo a critical-droplet flip there. Hence their spatially averaged concavity gap is zero. 

\medskip

Define the concavity gap $g(\omega)$ by
\begin{equation}
\label{eq:phi4}
g(\omega)=F_\omega\Big(\frac{u+v}{2}\Big) -\tfrac{1}{2}\Big[F_\omega(u)+F_\omega(v)\Big]\, .
\end{equation}
Then $g\ge 0$ almost everywhere and $g>0$ on the positive-measure set ${\mathcal S}$. Consequently,
\begin{equation}
\label{eq:phi5}
A_\phi\Bigl(\frac{u+v}{2}\Bigr)-\tfrac{1}{2}\Big[A_\phi(u)+A_\phi(v)\Big] =\int_\Omega g(\omega)\,\mu_{\widehat J}(d\omega)>0\, .
\end{equation}
The rest of the proof is identical to the final paragraph of the proof of Theorem~\ref{thm:finite}, with the full-measure set of conditionings supplied by Fubini chosen before selecting $[u,v]$.~$\diamond$

\medskip

{\bf Remark.}  The proof given here can also serve as an alternative proof of Theorem~4.10 in~\cite{NS26}.

\medskip

It is not difficult to use the results of Theorems~\ref{thm:subsetdense} and~\ref{thm:subsetnondense} to show that the possibility of $\sigma 2$ ground states is ruled out if ${\cal I}^<\subsetneq{\cal I}^>$.  Suppose that $J_{\rm max}<J_{\rm upper}$ and $J_{\rm min}=J_{\rm lower}$, so there's a gap between the upper ends of $I^>$ and $I^<$; suppose also that there is positive $\eta_J$-measure of $\sigma 2$ GSP's in $I^<$. As $J(b_0)$ is lowered from $J_{\rm upper}$, it must eventually reach the upper end of an interval $(u,v)\subseteq(J_{\rm min},J_{\rm max})$ with positive $\eta_J$-measure.  Either the $J_c^\sigma$'s are dense or not dense in a subinterval $(w,v)$ with $u\le w<v$, but both possibilities are ruled out by arguments identical to those in the proofs of Theorems~\ref{thm:subsetdense} and~\ref{thm:subsetnondense}.  A similar argument holds if $J_{\rm max}=J_{\rm upper}$ and $J_{\rm min}>J_{\rm lower}$.  We have therefore shown the following Corollary:

\medskip

\begin{corollary}
\label{cor:subset}
{\rm If either $J_{\rm max}<J_{\rm upper}$ or $J_{\rm min}>J_{\rm lower}$, the $\sigma 2$ property is absent from all GSP's in $\eta_J$ aside from a set of measure zero.}
\end{corollary}

We next turn to the remaining case when the intervals ${\cal I}^<$ and ${\cal I}^>$ are identical.

\subsubsection{$J_{\rm max}=J_{\rm upper}$ {\rm and} $J_{\rm min}=J_{\rm lower}$}
\label{subsubsec:nogap}

The remaining case is where ${\cal I}^<={\cal I}^>$ and the set ${\cal A}$ has the following properties: a) the set of $\sigma\in{\cal A}$, and their corresponding $J_c^\sigma$'s in $(J_{\rm min},J_{\rm max})$, is uncountable; b) each $\sigma\in{\cal A}$ has zero weight in~$\eta_J$; c) the full set of $\sigma\in{\cal A}$ has positive measure in~$\eta_J$; d) the $J_c^\sigma$'s are dense over {\it both\/} intervals~$(c,J_{\rm max})$ and $(J_{\rm min},d)$, for some $c$ and $d$ with $J_{\rm min}\le c<J_{\rm max}$ and $J_{\rm min}<d\le J_{\rm max}$, and e) the set of~$\sigma$'s with $J_c^\sigma$ in all open subintervals of  $(c,J_{\rm max})$ and $(J_{\rm min},d)$ have positive weight in~$\eta_J$.  As before, all $\sigma$'s referred to here and below have the $\sigma 2$ property.

\medskip

\begin{theorem}
\label{thm:nogap}
{\rm The event that the conditions above are all satisfied occurs with zero $\nu$-probability.}
\end{theorem}

\medskip

{\bf Proof.}  We will focus on the interval $(c,J_{\rm max})$. There is an immediate contradiction with Theorem~\ref{thm:flexconst} if all $\{\sigma:J_c^\sigma\in(c,J_{\rm max})\}$ are type~(i): simply lower $J(b_0)$ from slightly above $J_{\rm max}$ to $J_{\rm max}-\epsilon$, where $0<\epsilon<J_{\rm max}-c$. By the same arguments used in the proofs above, $\langle f\rangle_{\eta_J}$ must decrease.  If all $\{\sigma:J_c^\sigma\in(c,J_{\rm max})\}$ are type~(ii), then as before we need to consider the possible effect of $\sigma$'s with $J_c^\sigma<c$.  It is easily seen that if {\it all\/} $\sigma\in{\cal A}$ are type~(ii), then {\it any\/} decrease of $J(b_0)$ below $J_{\rm max}$ will increase $\langle f\rangle_{\eta_J}$, contradicting Theorem~\ref{thm:flexconst}.  If all $\{\sigma:J_c^\sigma\in(c,J_{\rm max})\}$ are type~(ii) only in this interval with $c>J_{\rm min}$, then the same argument as in the proof of Theorem~\ref{thm:subsetdense} leads to a contradiction with Theorem~\ref{thm:flexconst}. 

We therefore assume that for $J(b_0)\in(J_{\rm min},d)$ with $d\le c$ a positive fraction of $\sigma$'s are type (i) and a positive fraction are either type~(ii) or type~(iii) or both.  Compute ${\hat a}^+$ for these $\sigma$'s using the ``worst case'' scenario described in the proof of Theorem~\ref{thm:subsetdense}.   If $c+{\hat a}^+<J_{\rm max}$, then lower $J(b_0)$ from slightly above $J_{\rm max}$ to $c+{\hat a}^++\epsilon$, where $\epsilon>0$; as in the proof of Theorem~\ref{thm:subsetdense}, $\sigma$'s with $J_c<c$ will not contribute to the change in $\langle f\rangle_{\eta_J}$, which will therefore increase.

Suppose then that $c+{\hat a}^+=J_{\rm max}$ (it cannot exceed $J_{\rm max}$, which is now equal to $J_{\rm upper}$ which has the same property as a null point).  This is equivalent to the ``worst case'' scenario in the proof of Theorems~4.11 and~4.12 in~\cite{NS26} (see the discussion in the paragraph containing Eqs.~(11) and~(14) in~\cite{NS26}, respectively). The proof is then essentially identical to that of Theorems~4.11 and~4.12 in~\cite{NS26}.

There are two remaining possibilities: either there are no type (iii) bonds in $\{\sigma:J_c^\sigma\in(c,J_{\rm max})\}$ but a positive fraction of both type~(i) and type~(ii) bonds, or else there is a positive fraction of type~(iii) bonds for $\sigma$'s in this set.    Using the construction of the proofs of Theorems~4.11 and~4.12 in~\cite{NS26}, these situations are equivalent, and the remainder of the proof follows that of these theorems in~\cite{NS26}.~$\diamond$

\medskip

Combining the results of Theorems~\ref{thm:subsetdense},~\ref{thm:subsetnondense}, and~\ref{thm:nogap} leads to the following conclusion:

\medskip

\begin{corollary}
\label{cor:final}
{\rm Given a translation-covariant zero-temperature metastate~$\eta_J$, the set of GSP's in its support with the $\sigma 2$ property has zero $\eta_J$-measure.}
\end{corollary}

\section{A necessary condition for ground state multiplicity}
\label{sec:necessary}

As noted in Sect.~\ref{sec:intro}, Theorem 9.3 of~\cite{NS22} shows that the presence of either kind of $\sigma$-criticality is a sufficient (but not necessary) condition for the presence of multiple ground state pairs in the EA~spin glass. Ruling out these sufficient conditions naturally leads to the question of finding a rigorous {\it necessary\/} condition for GSP multiplicity to occur. An earlier conjecture (Claim 9.5 in~\cite{NS22}) connected ground state multiplicity to the average number of critical droplet boundaries that include a fixed bond in a given GSP. In this section, using ideas from the nonrigorous argument in~\cite{NS22}, we provide a rigorous argument to establish a necessary condition.

\medskip

{\bf Definition~\cite{NS22}}.  Consider an infinite sequence of volumes $\Lambda_L$, with $L\to\infty$. If, in a positive fraction of these volumes, a change in boundary condition from periodic to antiperiodic in one coordinate direction changes the spin configuration inside any (large, but small compared to $\Lambda_L$) fixed window centered at the origin, then we will say that the zero-temperature periodic boundary condition metastate~$\kappa_J$ is supported on observably multiple ground state pairs. 

\medskip

{\bf Remark}. It was proved in~\cite{NS01c} that the distinct ground state pairs generated in this way are necessarily mutually incongruent. In the rest of the paper  ``multiple ground state pairs'' will mean observably multiple.

\medskip

Consider a cube~$\Lambda_L$ of side $L$ in $d$ dimensions, centered at the origin with periodic boundary conditions, and let $\sigma^{(L)}$ denote the corresponding GSP in $\Lambda_L$. Define
\begin{equation}
\label{eq:K}
K_L^{(b)} =\sum_{e\in E_L} {\mathbf 1}_{b^*\in\partial D(e,\sigma^{(L)})}\, ,
\end{equation}
that is, \(K_L^{(b)}\) specifies the number of finite-volume critical-droplet boundaries that include the dual \(b^*\) of the bond $b$ in $\sigma^{(L)}$.  

\medskip

\begin{theorem}
\label{thm:necessary}
{\rm Let} ${\mathbb E}[\cdot]$ {\rm denote an average over couplings.  Suppose that, for a fixed observation window} $W$ {\rm and a deterministic volume sequence} $L_n$, {\rm changing periodic to antiperiodic boundary conditions in a fixed coordinate direction changes at least one bond product in} $W$ {\rm in a positive fraction of these volumes on a set of coupling realizations of positive probability. Then for every fixed bond} $b$, $\limsup_{n\to\infty}{\mathbb E}[K_{L_n}^{(b)}]/L_n > 0$. {\rm In particular,} $\limsup_{L\to\infty}{\mathbb E}[K_L^{(b)}]=\infty$.
\end{theorem}

\medskip

{\bf Proof.}  Let $W=\Lambda_{L_0}$ with $L_0$ fixed as $L\to\infty$. Choose a coordinate direction exhibiting the observable boundary-condition response. There are \(L\) parallel seams \(S_j\) on which reversing all coupling signs implements an antiperiodic twist. (Here `seam' denotes a set of bonds crossing one cut of the torus, perpendicular to a chosen coordinate direction, so that reversing the signs of all couplings on that set changes periodic to antiperiodic boundary conditions in that direction.)

Let \(A_L^{(j)}\) denote the event that reversing all coupling signs along \(S_j\) changes at least one bond product in \(E_W\), where $E_W$ is the edge set in $W$. Order the seam bonds as \(e_1,\ldots,e_{L^{d-1}}\), and let \(J^{(i-1)}\) be \(J\) with the signs of \(e_1,\ldots,e_{i-1}\) reversed. Define $B_e(J)$ as the event that ${\partial D(e,\sigma^{(L)}(J))\cap E_W\ne\emptyset}$.  If the full seam reversal changes a bond product in \(W\), at least one step does. Specifically, with all other couplings fixed, the minimizing configurations subject to the two possible values of the bond product at $e_i$ do not depend on $J(e_i)$. Thus reversing $J(e_i)$ either leaves the ground state unchanged or flips its critical droplet. A change of a bond product in $W$ at this step therefore implies $B_{e_i}(J^{i-1})$.  A union bound then gives 
\begin{equation}
\label{eq:B}
\mathbb P(A_L^{(j)}) \le \sum_{i=1}^{L^{d-1}} \mathbb P\left(B_{e_i}(J^{(i-1)})\right)\, .
\end{equation}
Note that a seam that does not intersect $W$ can be moved without changing $\sigma_x^{(L)}\sigma_y^{(L)}$ with $\langle x,y\rangle\in W$; i.e., if one applies antiperiodic BC's along \(S_j\) in one trial and along \(S_k\) in another trial, and both seams avoid \(W\), then $\sigma_x^{(L)}\sigma_y^{(L)}$ is the same in the two resulting ground state pairs; consequently $A_L^{(j)}=A_L^{(k)}$.

For each seam, the sequential-flip union bound~(\ref{eq:B}) applies. Reverse its couplings in a deterministic order. A net change of $\sigma_x^{(L)}\sigma_y^{(L)}$ in \(W\) requires a change at some step. Since any deterministic collection of coupling-sign reversals preserves the disorder law, $J^{(i-1)}\stackrel{\mathrm d}=J\Rightarrow{\mathbb P}\left(B_{e_i}(J^{(i-1)})\right)={\mathbb P}\left(B_{e_i}(J)\right)$, so
\begin{equation}
\label{eq:reversal1}
{\mathbb P}(A_L^{(j)}) \le \sum_{i=1}^{L^{d-1}}\mathbb P(B_{e_i}(J))
\end{equation}
and therefore
\begin{equation}
\label{eq:reversal2}
{\mathbb P}(A_L^{(j)}) \le \sum_{e\in S_j} \mathbb P\Bigl[\bigl(\partial D(e,\sigma^{(L)})\cap E_W\bigr)\ne\emptyset\Bigr]\, .
\end{equation}

Now average over the \(L-O(L_0)\) seams avoiding \(W\), and let \(A_L\) denote the common event \(A_L^{(j)}\) for seams avoiding \(W\). Note that $\sum_j\sum_{e\in S_j} \mathbf1_{{\partial D(e,\sigma^{(L)})\cap E_W\ne\emptyset}} \le \sum_{b\in E_W}K_L^{(b)}$. The seams are disjoint subsets of the torus edge set, so
\begin{equation}
\label{eq:torus}
\mathbb P(A_L) \le \frac{1}{\bigl(L-O(L_0)\bigr)}\sum_{b\in E_W}\mathbb E[K_L^{(b)}]\, .
\end{equation}

Now fix the deterministic volume sequence $L_n$ used in the definition of observable multiplicity, and set $F_N(J) = N^{-1}\sum_{n=1}^N{\bf 1}_{A_{L_n}}(J)$. Observable multiplicity implies that $\limsup_NF_N(J) > 0$ on a set of coupling realizations of positive probability. Since $0\le F_N\le 1$, reverse Fatou gives $\limsup_N{\mathbb E}[F_N]\ge{\mathbb E}[\limsup_NF_N]>0$. Consequently $\limsup_n{\mathbb P}(A_{L_n}))>0$.  Therefore observable multiplicity, in the sense used in the definition above, forces 
\begin{equation}
\label{eq:limsup}
\limsup_{L\to\infty} \frac1L\sum_{b\in E_W}\mathbb E[K_L^{(b)}]>0
\end{equation}
along a sequence with nonvanishing boundary-condition response. The expected number of bonds whose critical-droplet boundaries contain \(b^*\) must grow at least linearly along that sequence.  Under the i.i.d.~disorder law, translations and coordinate symmetries of the periodic torus give $\mathbb E[K_L^{(b)}]=a_L$ for every bond \(b\). Thus (\ref{eq:torus}) yields 
\begin{equation}
\label{eq:last}
\frac{a_L}{L} \ge \frac{1-O(L_0/L)}{|E_W|}\mathbb P(A_L)
\end{equation}
which implies $\limsup_{L\to\infty}\frac{\mathbb E[K_L^{(b)}]}{L}>0$ for every fixed bond \(b\).  Taking $L = L_n$ in (\ref{eq:last}) and using $\limsup_{n\to\infty}{\mathbb P}(A_{L_n})>0$ gives $\limsup_{n\to\infty} {\mathbb E}[K_{L_n}^{(b)}]/L_n > 0$ for every fixed bond $b$.~$\diamond$

\bigskip

{\bf Acknowledgment.}  This work was initiated at the Aspen Center for Physics, which is supported by National Science Foundation grant PHY-2210452.

\bibliography{refs.bib}
\end{document}